\documentclass[a4paper]{article}

\usepackage{INTERSPEECH2022}
\usepackage{xcolor}
\usepackage{url}
\usepackage{comment}

\usepackage{pifont}
\newcommand{\cmark}{\ding{51}}%
\newcommand{\xmark}{\ding{55}}%

\usepackage{tikz}
\usetikzlibrary{automata, arrows.meta, positioning}

\title{BUT CHiME-7 system description}
\name{Martin Karafiát$^1$, Karel Veselý$^1$, Igor Szőke$^1$, Ladislav Mošner$^1$, Karel Beneš$^1$, Marcin Witkowski$^2$, Germán Barchi$^3$, Leonardo Pepino$^3$}
\address{
  $^1$Brno University of Technology\\
  $^2$AGH University of Krakow\\
  $^3$University of Buenos Aires}
\email{karafiat@fit.vutbr.cz, iveselyk@fit.vutbr.cz, szoke@fit.vutbr.cz, imosner@fit.vutbr.cz, ibenes@fit.vutbr.cz, witkow@agh.edu.pl, gbarchi@dc.uba.ar, lpepino@dc.uba.ar}

\begin{document}

\maketitle
\begin{abstract}
  This paper describes the joint effort of Brno University of Technology (BUT), AGH University of Krakow and University of Buenos Aires on the development of Automatic Speech Recognition systems for the CHiME-7 Challenge.
  We train and evaluate various end-to-end models with several toolkits.
We heavily relied on Guided Source Separation (GSS) to convert multi-channel audio to single channel. The ASR is leveraging speech representations from models pre-trained by self-supervised learning, and we do a fusion of several ASR systems.
In addition, we modified external data from the LibriSpeech corpus to become a close domain and added it to the training.

Our efforts were focused on the far-field acoustic robustness sub-track of Task 1 - Distant Automatic Speech Recognition (DASR), our systems use oracle segmentation.

\end{abstract}

\section{Introduction}

This paper describes the BUT Automatic Speech Recognition (ASR) system for the CHiME-7 Speech to Text Transcription (STT) Challenge. We present the detailed description of the datasets, as well as technical details for the development of ASR subsystems and the fusion.


\section{Data}
\label{sec:data}

Our training setup is derived from the released baseline ESPnet~\cite{watanabe2018espnet} recipe\,\footnote{https://github.com/espnet/espnet/tree/master/egs2/chime7\_task1/asr1}. The training setup is composed from {\em Chime6}~\cite{chime6}, {\em Mixer6} (LDC2013S03), and {\em Dipco}~\cite{dipco} datasets, where {\em Chime6} and {\em Mixer6} have training/dev/eval data split, and {\em Dipco} is used only for development and evaluation.

Table~\ref{tab:data} describes all the training data selection setups used in our system building. The {\em baseline} and {\em baseline + mixer6gss} shows amount of training data used in baseline recipe before and after fixing bug in baseline recipe (missing GSS for {\em Mixer6} data). The target data are always processed with GSS, and the amount of GSS-processed training data is very limited in the baseline recipe. Therefore, we decided to modify the data preparation setup, and we created new data sets {\em limited} and {\em limited+libri} with following procedure:
\begin{itemize}
    \item We added GSS-enhanced Mixer6 training data (see section~\ref{gss}), which were missing due to a bug in the baseline recipe.
    \item 3-way speed perturbation is applied only to the GSS data.
    \item The data were re-balanced to increase the weight of the GSS-enhanced data. We randomly chose 80 hours of each combination of \verb|(Chime6, Mixer6) x (IHM, MDM)| type of the training utterances.
    \item LibriSpeech~\cite{librispeech} data augmented with {\em background speaker} were added.
\end{itemize}

The {\em limited} dataset in Table~\ref{tab:data} consists of the speed-perturbed `full amount' of the GSS-processed data and the subsampled IHM+MDM data. The {\em limited+libri} has the augmented LibriSpeech added, and the {\em gss-only} consists purely of speed-perturbed GSS-processed data ({\em Chime6}, {\em Mixer6}).

\begin{table}[tb]
  \caption{Training data selections}
  \vspace{-2mm}
  \label{tab:data}
  \centering
  \begin{tabular}{ r | r | c }
    \toprule
    \textbf{Dataset}           & \textbf{\# hours} & \textbf{GSS part}  \\
    \midrule
    {\em baseline}             &  5922             & 1.7\% \\
    {\em baseline + mixer6gss} &  6301             & 2.8\% \\
    {\em limited}              &  611              & 29.1\% \\
    {\em limited+libri}        &  1108             & 16.1\% \\
    {\em gss-only}             &  288              & 100\% \\
    \bottomrule
  \end{tabular}
  \vspace{-4mm}
\end{table}

\subsection{Multispeaker augmentation}

\begin{table}[b]
  \vspace{-4mm}
  \caption{Used RIRs}
  \vspace{-2mm}
  \label{tab:rirs}
  \centering
  \begin{tabular}{ r | r }
    \toprule
    \textbf{Dataset}            & \textbf{\#RIRs}  \\
    \midrule
    AIR14~\cite{Jeub2009}       & $214$             \\
    REVERB~\cite{Kinoshita2013} & $192$               \\
    RWCP~\cite{Nakamura2000}    & $3240$       \\
    ReverbDB~\cite{ReverbDB}    & $1364$              \\
    Synthetic                   & $10000$              \\
    \bottomrule
  \end{tabular}
\end{table}

\label{augmentation}
We used LibriSpeech dataset to simulate `multispeaker' condition. All recordings were augmented by another randomly selected recording, coming from the same subset of LibriSpeech, to insert a so called `background speaker'. The background speaker was expanded by 4 seconds of silence on both sides. Then, this expanded audio was merged with the original audio file starting in random position (and the expanded audio was looped if the end of the background speaker was achieved sooner than the end of the original audio). We maintain the Signal-to-Noise Ratio (SNR) between these two audios in the range of 5dB - 12dB randomly. Each of the audio was reverberated by a single Room Impulse Response (RIR) randomly selected from a pool shown in Table~\ref{tab:rirs}. Finally,  one of the following codecs (MP3, AMR, AMR-WB, G.711, G.726, G.729, GSM-FR, TETRA, GSM-EFR) was applied with a probability of 1/7 on the resulting merged audio. The resulted merged audio sounds like the original LibriSpeech in a reverberated environment with random speaker speaking in the background.
 Synthetic RIRs were generated using the image method~\cite{Habets2010}. We randomly sampled the dimensions of the room as $width=[1.5-5.5]$, $height=[2.0-9.5]$, $length=[2.5-16.5]$. The source and microphone were randomly placed in the room and the wall reflections were anything between $\beta=[0.45-0.95]$. We used only a `visible' subset of RIRs from ReverbDB~\cite{ReverbDB}.

\section{Speech Enhancement}

\subsection{Non-neural approach - GSS}
\label{gss}

The speech enhancement was based on the baseline system provided by the organizers.
First, $k = 80\%$ channels were selected for further processing using the Envelope Variance~\cite{wolf2014channel} method. The baseline Guided Source Separation (GSS)~\cite{raj2022gpu} was used as a primary speech enhancement technique, which integrates Weighted Prediction Error (WPE) dereverberation method ~\cite{yoshioka2012generalization}, estimation of target and undesired component time-frequency masks using oracle diarization output and Complex Angular Central Gaussian Mixture Model~\cite{ito2016complex} posteriors, and the mask-based Minimum-Variance Distortionless Response (MVDR) beamformer~\cite{souden2009optimal}. The GPU-based implementation of GSS~\cite{raj2022gpu} with its default parameters was used in the experiments as a baseline.
Table~\ref{tab:gss_se_dev} contains the Word Error Rates (WER) results obtained using different non-neural speech enhancement methods for the pretrained ASR model on the development sets used in the Challenge. Note, the pretrained models was WAVLM based model trained with baseline recipe and uploaded by organizers. The running script was simply called with {\scriptsize\verb|--use-pretrained popcornell/chime7_task1_asr1_baseline|} option to quickly analyze speech enhancement technique.



\begin{figure} [!h]
    \centering
    \includegraphics[width=0.8\linewidth]{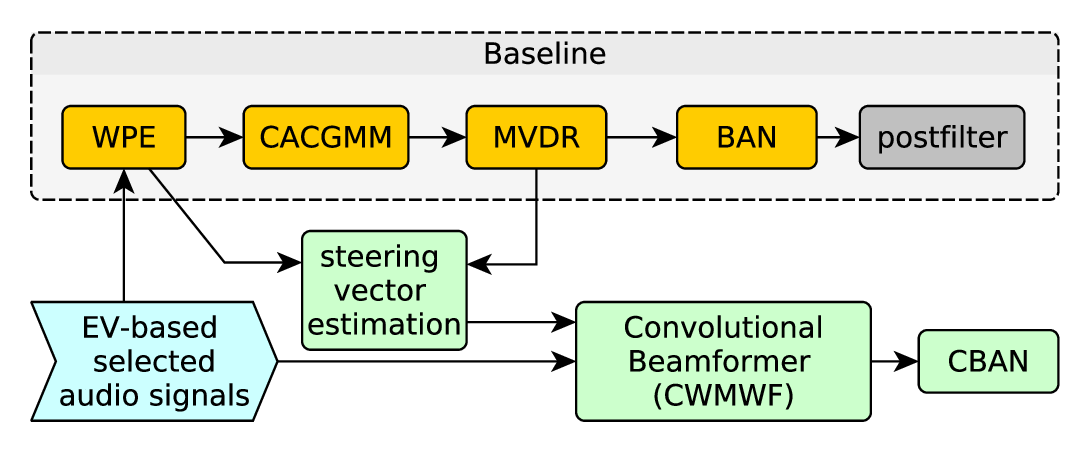}
    \caption{Evaluated non-neural speech enhancement.}
    \label{fig:preprocessing}
\end{figure}

To validate the performance of stronger filtering, we experimented with reuse of CACGMM estimated target masks as a post-filter applied to the output of the beamformer. This result is presented in the second row of Table \ref{tab:gss_se_dev}. To investigate the performance of using the filter that allows for milder attenuation than using masks directly, we have also experimented with the use of a Convolutional Weighted Multichannel Wiener filter (CWMWF)~\cite{fras2022convolutive} as a convolutional beamformer which performs dereverberation and source separation jointly. The general diagram of the investigated systems is depicted in Figure~\ref{fig:preprocessing}.

A steering vector was selected as the principal eigenvector of the spatial covariance matrix (SCM) calculated using the multiplication of the multichannel output of WPE and the estimated MVDR beamformer from the baseline system. This steering vector and multichannel input were used to estimate CWMWF. A Convolutional Blind Analytical Normalization (CBAN) was applied to the output signal to compensate for attenuation introduced by the beamformer. The weighting factors for a single frequency bin $k$ were calculated similarly to non-convolutional Blind Analytical Normalization (BAN)~\cite{warsitz2007blind} with CWMWF coefficients $\textbf{w}(k) \in \mathbb{C}^{IL}$ and convolutional noise covariance matrix $\boldsymbol{\Phi}_N(k) \in \mathbb{C} ^ {IL \times IL}$, where $I, L$ are the number of channels and the number of filter taps, respectively. For audio segments in which the maximum signal amplitude after analytical normalization exceeded one, additional peak normalization was applied to prevent clipping.

In the baseline system, the multichannel signal is extended by the left and right context of 15 seconds by default to compute WPE dereverberation filter and estimate masks with CACGMM. Then the context is dropped when calculating MVDR beamformer coefficients. Since the convolutional filter performs joint dereverberation and separation the context for this filter has been greatly reduced to minimize artifacts resulting from estimation filter from longer segments. So far in the experiments we have shown that after reduction of the context to 1 second the CWMWF achieves comparable results to the baseline system. Note that in those experiments, ASR system was not fine-tuned for the new domain of audio data introduced by stronger filtering. Therefore, we claim that presentation of the processed data to the ASR system is crucial.

\begin{table}
  \vspace{-4mm}
  \caption{WER [\%] using the pretrained baseline ASR with different non-neural speech enhancement on development sets.}
  \vspace{-2mm}
  \begin{tabular}{lrrr}
    \toprule
     Method                   & Chime6  & Dipco  & Mixer6   \\
    \midrule
    GSS (the baseline)      &  33.3  &  34.6  & 21.8  \\
    GSS (baseline+postfilter) &  35.8  &  36.8  & 23.5  \\
    GSS (context 1s)        & 47.4   & 56.8   &  24.6\\  
    GSS+CWMWF+CBAN            &  49.5  &  58.4 & 26.9  \\
    \bottomrule
  \end{tabular}
  \label{tab:gss_se_dev}
\end{table}




\subsection{Neural approach}
\label{NN_enhancement}

\begin{figure}[t]
  \centering
  \includegraphics[width=0.95\linewidth]{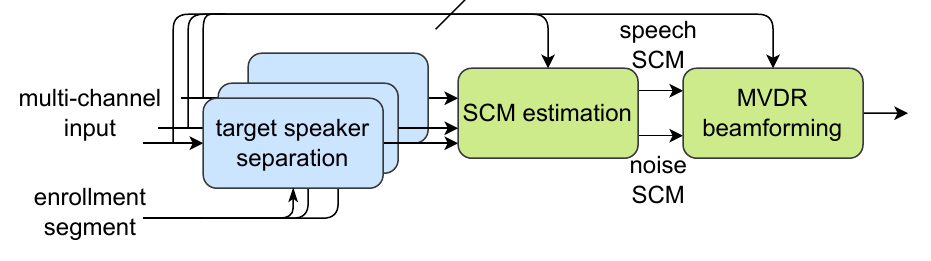}
  \vspace{-2mm}
  \caption{Multi-channel speech enhancement based on neural networks.}
  \label{fig:nn_se}
\end{figure}

Contrary to GSS-based pre-processing, we experimented with discriminative models. A general structure of the approach is depicted in Figure \ref{fig:nn_se}. In the first stage, we employ Target Speaker Extraction (TSE) models to provide per-channel estimates of the speaker-of-interest speech where the enrollment utterances are provided by the oracle diarization. Given the predictions, input audio, and the assumption that speech and noise are not correlated, we estimate noise by subtracting speech from mixtures. Subsequently, both speech and noise signals are transformed via STFT to provide ratio masks. A speech mask is computed as a ratio between the power spectrum of predicted speech and the sum of the power spectra of speech and noise. A noise mask is obtained analogically. We follow a standard approach to estimate spatial covariance matrices of speech and noise using the masks \cite{Heymann2016-nn-bf}. They are used to compute beamforming weights following Minimum Variance Distortionless Response (MVDR) approach \cite{Capon_MVDR, Souden2010_MVDR_form}, which combine input channels previously selected analogically to the baseline GSS.

We experimented with three types of models for TSE --- SpeakerBeam \cite{Zmolikova2019-speakerbeam}, DPCCN \cite{Han2022-dpccn}, and a down-scaled version of TF-GridNet \cite{Wang2023-tfgridnet} conditioned on the enrollment utterance through Feature-wise Linear Modulation (FiLM) \cite{Perez2018-film}. The TF-GridNet-based TSE was inspired by that used in iNeuBe-X \cite{Cornell2023-ineubex}. Therefore, we also use the encoder and TCN modules of TCNDenseNet \cite{Wang2021-tcndensenet} to extract speaker embedding from the enrollment segments. Compared to TSE in iNeuBe-X, multiple channels are not concatenated to form the input, but for consistency with other explored models, each channel is processed independently. For computation reasons, we set number of blocks B=5 and LSTM hidden units H=128 in TF-GridNet. The number of repeats in TCNDenseNet was reduced to 2. For SpeakerBeam and DPCCN, we used hyperparameters following the respective papers.

All the TSE networks were trained with the SNR objective. Since the reference speech signals are required during training, we dynamically simulated the data. As a source of speech, we used the train-clean-360 subset of Librispeech. As a source of background noise, MUSAN and WHAM \cite{Wichern2019-wham} datasets were utilized. The SNR intervals for adding noise are shown in Table \ref{tab:neural_se_dev}. To simulate room acoustic conditions, we reverberated the signals with simulated room impulse responses (with RT60 from 0.3 to 0.9) and those drawn from the ReverbDB corpus \cite{ReverbDB}. In addition to the target speaker, audio mixtures contain up to 2 other voices, while only the target speaker's speech examples are also allowed.

The comparison of the TSE methods on the development sets is presented in Table \ref{tab:neural_se_dev}, this is with the baseline ASR system employed. We observed noticeable degradation for Chime6 and Dipco compared to the baseline GSS. We hypothesize it is mainly caused by the domain mismatch of test and training data and the inability of neural models to generalize. While poor generalization of time-domain models is common, we did not observe improvements from using TF-GridNet. However, we can confirm that TF-GridNet provides better predictions compared to other models as we observed differences in results when experimenting with the approach where SCMs were computed directly using network outputs (without masking).


\begin{table}[t]
    \caption{WER [\%] using the pretrained baseline ASR with different neural speech enhancement on development sets.}
    \vspace{-2mm}
    \centering
  \begin{tabular}{lcrrr}
    \toprule
     Method                  & Mixing SNR & Chime6  & Dipco  & Mixer6   \\
    \midrule
    SpeakerBeam & 5 -- 30 & 54.4 & 50.4 & 21.6 \\
    DPCCN & 5 -- 30 & 50.9  & 48.6 & 21.6 \\
    TF-GridNet & 5 -- 30 & 50.9 & 49.8 & 21.8\\
    \midrule
    SpeakerBeam & $-$4 -- $-$10 & 52.6 & 47.4 & 21.3 \\
    \bottomrule
  \end{tabular}
  \label{tab:neural_se_dev}
\end{table}

\section{Automatic Speech Recognition systems}

We experimented with Encoder-Decoder ASR models coming from various toolkits.
Some ASR systems are built on top of `fixed' pre-trained feature representations,
while in other cases these pre-trained models are `fine-tuned'.


\subsection{HuggingFace}
We fine-tuned the WavLM large model by adding a conformer layer and a RNN transducer ~\cite{graves2012rnnt} on top of it. The conformer layer consists of 4 attention heads, a kernel size of 31, downsampling in time by a factor of 4 and dropout with rate 0.1. These are the default values of the torchaudio implementation we used. The predictor network consists of 2 LSTM layers and we also used the RNNT loss from torchaudio implementation. The model was fine-tuned using GSS data from Chime6 and Mixer6 datasets (i.e. {\em gss-only} in Table \ref{tab:data}). The CNN encoder of the WavLM model was frozen during the fine-tuning, updating only the weights from the transformer layers and the RNNT modules. We applied a learning rate of 1e-4 with a warmup of 1000 steps, and a batch size of 16 samples. The model was trained for 30000 steps using the AdamW optimizer with a weight decay of 0.005. For the generation of the transcriptions, we used the RNNT beam search decoder from torchaudio with a beam size of 20.


\subsection{WavLM with speaker information}

We also explored adding the target speaker information to the WavLM model. This could potentially help the model to focus on transcribing the target speaker when the GSS signal contains interference from non-target speakers. We used x-vectors from VBx as speaker embeddings, linearly projected them and summed them to the input of the first transformer layer. We learned a scale factor for this projected embedding, which was initialized to 0 so that at the beginning of training, no speaker information is added and the model is equivalent to the original WavLM. We verified that during training, the scale factor increased in magnitude, suggesting that the model is in fact using the target speaker information. In order to train the model with a larger amount of speakers, we augmented the training data with artificial mixtures of up to 3 speakers from Librispeech, resulting in around 200k extra utterances. 

\subsection{EspNET}
We selected two pre-trained models via S3PRL \verb|wav2vec2_large_lv60_cv_swbd_fsh|~\cite{baevski:w2vec} and \verb|wavlm_large|~\cite{chen2022wavlm} for building systems with ESPnet toolkit.

\vspace{1mm}
\textbf{WavLM Large + Transfomer}\quad
This system followed the baseline recipe, therefore the initial {\em wavlm\_large} model
is followed by Transformer based model: encoder (12 layers) -- decoder (6 layers).

\textbf{Wav2Vec2 + Conformer}\quad
This system is based on the Conformer architecture~\cite{gulati2020conformer} and is composed of 12 encoder layers and 6 decoder layers. The conformer encoder layer incorporates, in addition to a self-attention module, a convolutional layer in between of two feed-forward modules. The decoder was built using masked self-attention as well as cross-attention between the encoder embeddings and the decoder. Each encoder and decoder layer outputs 512 dimensional embeddings; attention is done with 8 parallel heads and the feed-forward module expands the data into 2048 dimensions. We used the standard ESPnet2 training recipe with 40k warm-up steps and learning rate $2\cdot 10^{-3}$.

Both models were trained with frozen update of the pre-trained models (WavLM, Wav2Vec2).
The model output are 500 Sentencepiece \cite{sentencepiece} unigram units. The model is trained with the joint CTC/Attention loss with the CTC weight of 0.3.

The systems were further fine-tuned with the {\it gss-only} data (in Table \ref{tab:data}) using lower learning rate, for this step we `defreezed' weights in the pre-trained models (WavLM, Wav2Vec2) from S3PRL.

\ifx{\TABLE_IS_DISABLED}
\begin{table*}[t]
  \caption{Results}
  \label{tab:results}
  \centering
  \vspace{-2mm}
  \begin{tabular}{@{}cll|cccc@{}}
    \toprule
        &              &                  &       & \multicolumn{3}{c}{Dev {[}\% WER{]}} \\
        & Arch.        & Train. set       & Enh.  & Chime6 & Dipco & Mixer6  \\
    \midrule
    1   & WavLM Large + Transformer-Transformer (ESPnet, fixed feat.)   &  limited          & GSS   & 32.2   & 33.2  & 21.1    \\
    2   & WavLM Large + Transformer-Transformer (ESPnet, fixed feat.)   &  limited+libri    & GSS   & 30.2   & 31.9  & 19.9    \\
    3 (S) & WavLM Large + Transformer-Transformer (ESPnet, fine-tuned feat.)  &   (2) $\rightarrow$gss-only & GSS   & {\bf 25.4}  &  {\bf 29.8}  &  18.7 \\
    4   & Wav2Vec2 Large + Conformer-Transformer (ESPnet, fixed feat.)   &  limited+libri    & GSS  & 37.9  & 40.3   & 21.7 \\
    5  & Wav2Vec2 Large + Conformer-Transformer (ESPnet, fine-tuned feat.)   &   (4) $\rightarrow$gss-only & GSS   &  31.3     &  38.2   & 20.2 \\
    6   & WavLM Large + CTC (HuggingFace, fine-tuned feat.)        & gss-only & GSS & 40.2 & 50.4 & 28.7 \\ 
    7   & WavLM Large + Conformer-Transducer (HuggingFace, fine-tuned feat.) & gss-only & GSS & 28.2 & 36.2 & {\bf 17.2} \\ 
    8   & WavLM Base Plus + Zipformer-Transducer (K2, fixed feat.) & limited  & GSS & 40.9 & 43.6 & 24.9 \\
    \midrule
    F1  & 3 + 7       & & GSS & 25.3 & 31.2 & 18.3  \\
    F2  & 3 + 7 + 8   & & GSS & ?    & ?    & ?     \\
    \bottomrule
  \end{tabular}
\end{table*}
\fi

\begin{table*}[ht]
  \label{tab:results}
  \centering
  \vspace{-2mm}
  \captionof{table}{Word Error Rates [\%] obtained on \textit{dev}
    parts of the datasets for various system architectures and
    training data.}\label{tab:results}
    \addtolength{\tabcolsep}{-4pt}
    \begin{tabular}{@{}c|lc|l|lc|ccc@{}}
    \toprule
        & Pre-trained model & fine-tuned \, &  Architecture (toolkit)
    & Training data       & Enh. & Chime6 & Dipco & Mixer6  \\
    \midrule
    0   & WavLM Large & \xmark & Transformer-Transformer (ESPnet)   &
    \textit{baseline}          & GSS   & 33.5   & 35.4  & 23.7    \\
    1   & WavLM Large & \xmark & Transformer-Transformer (ESPnet)   &
    \textit{limited}          & GSS   & 32.2   & 33.2  & 21.1    \\
    2$^{\dagger}$   & WavLM Large & \xmark & Transformer-Transformer
    (ESPnet)   & \textit{limited+libri}    & GSS   & 30.2   & 31.9  & 19.9
    \\
    2(S) & WavLM Large & \cmark & Transformer-Transformer (ESPnet)  &
    (2$^{\dagger}$) $\rightarrow$\textit{gss-only} & GSS   & {\bf 25.4}  &
    {\bf 29.8}  &  18.7 \\ \midrule
    3$^{\dagger}$   & WavLM Large & \xmark & Transformer-Transformer
    (ESPnet)   &  \textit{limited+libri}    & GSS\_postfilter  & 32.9   & 33.5
    & 21.8    \\
    3  & WavLM Large & \cmark & Transformer-Transformer (ESPnet)  &
    (3$^{\dagger}$) $\rightarrow$gss-only & GSS\_postfilter   &  26.6
    &  31.1  &  19.6 \\ \midrule
    4$^{\dagger}$ & WavLM Large & \cmark & Conformer-Transformer
    (ESPnet)  &   \textit{limited+libri} & GSS   & 29.2  &  30.9  &   17.4 \\
    4     & WavLM Large & \cmark & Conformer-Transformer (ESPnet)  &
    (4$^{\dagger}$) $\rightarrow$gss-only & GSS   & 25.7  &  30.1  &
    16.9 \\
    5$^{\dagger}$  & Wav2Vec2 Large & \xmark & Conformer-Transformer
    (ESPnet)   &  \textit{limited+libri}    & GSS  & 37.9  & 40.3   & 21.7 \\
    5   & Wav2Vec2 Large & \cmark & Conformer-Transformer (ESPnet)   &
    (5$^{\dagger}$) $\rightarrow$\textit{gss-only} & GSS   &  31.3     &  38.2
    & 20.2 \\ \midrule
    6   & WavLM Large & \cmark & CTC (HuggingFace)        & \textit{gss-only} &
    GSS & 40.2 & 50.4 & 28.7 \\ 
    7   & WavLM Large & \cmark & Conformer-Transducer (HuggingFace) &
  \textit{gss-only} & GSS & 28.0 & 36.0 & 17.2 \\ 
    8    & WavLM Large & \cmark & Conformer-Transducer (HuggingFace) &
    \textit{gss-only} & GSS & 26.3 & 31.7 & {\bf 15.8} \\
    9   & WavLM Large & \cmark & LinSum XVector & \textit{librimix+gss} & GSS &
    28.4 & 34.1 & 17.8 \\
    \midrule
    10  & WavLM Large & \xmark & Zipformer-Transducer (K2) & \textit{limited} & GSS &
    30.0 & 32.5 & 17.5 \\
    \midrule
    F1  &  -- & & 2 + 8    & -- & GSS & 25.3 & 31.2 & 18.3  \\
    F2  & --  & & 4 + 8 
    & -- & GSS & 25.0 & 30.2 & 16.0  \\
    F3  &  -- & & 3 + 4 + 8    & -- &  -- & {\bf 23.8} & {\bf 28.8} &
    {\bf 15.4}  \\
    \bottomrule
  \end{tabular}
  \vspace{-2mm}
\end{table*}

\subsection{K2}
The K2 codebase was extended to accomodate S3PRL models, such as {\em wavlm-large} \cite{chen2022wavlm}, as a fixed feature extraction. The {\em wavlm-large} model transforms raw audio signal to 1024 dimensional embeddings with 20ms time-shift. We duplicate each embedding vector to create a stream with 10 ms time-steps.

On top of these embeddings, we trained a {\em Zipformer-Transducer} model from the streaming ASR recipe\,\footnote{\url{https://github.com/vesis84/icefall/tree/master/egs/librispeech/ASR/pruned_transducer_stateless7_streaming}}. Due to using a pre-encoder, we removed the \verb|Conv2dSubsampling| front-end module in Zipformer and we reduced the numbers of \verb|ZipformerEncoderLayer| modules to : \verb|[1,1,1,1,1]|.
To link the {\em wavlm-large} pre-encoder to {\em Zipformer encoder}, a trainable linear transform without bias was introduced to reduce the dimensionality from 1024 to 384 dimensions. The model is using a {\em stateless} Predictor network in Transducer architecture~\cite{stateless_rnnt}. And, the training is accelerated by a {\em pruning} secondary output-layer that pre-selects the candidate tokens~\cite{pruned_rnnt}. The Zipformer-Transducer model uses half-precision, the model size is 386 MB, and the output are 500 sentence-piece \cite{sentencepiece} unigram units.
The {\em wavlm} pre-econder had frozen parameters.

We trained for 15 epochs with base learning-rate 0.025, and we observed an increased level of overfitting. The overfitting was apparent for the \verb|valid_pruned_loss| objective, starting after the 4th epoch. We used the {\em limited} dataset from Table~\ref{tab:data}. We did not use SpecAugment because of using the S3PRL feature transform pre-encoder.

For decoding we used the \verb|fast_beam_search_nbest| method with default hyper-parameters. For the final fusion, we exported $N$-best lists of size up to 200 generated by sampling a lattice.



\subsection{System fusion}
To facilitate effective fusion of outputs of the different systems, we first compact each resulting $N$-best list into a CTM using the Hystoc tool~\cite{hystoc} with the temperature parameter set to 1.0.
To merge the CTMs, we used NIST Rover~\cite{rover}, where the selection of output words is done according to the word frequency and maximum confidence.
In Rover, we tuned the $\alpha$ parameter, which is a trade-off between frequency of word occurrence and maximum word confidence, as well as the null word confidence (also known as blank symbol confidence).
In specific, $\alpha$ and null word confidence were set to 0.8 and 0.4, respectively (for the best performing fusion F3).
We did not use the time information during fusion.

\begin{minipage}[ht]{0.8\linewidth}
    \vspace{7mm}
    \begin{tikzpicture} [node distance = 4cm, on grid, auto, transform canvas={scale=0.57}]
        \node (q0) [state,inner sep=15pt,minimum size=0pt, initial
          text = {}] {};
        \node (q1) [state,inner sep=15pt,minimum size=0pt, right = of
          q0] {};
        \node (q2) [state,inner sep=15pt,minimum size=0pt, right = of
          q1] {};
        \node (q3) [state,inner sep=15pt,minimum size=0pt, right = of
          q2, accepting] {};

        \path [-stealth, line width=5pt]
            (q0) edge node {A / 1.0} (q1)
            ;

        \path [-stealth, line width=5pt]
            (q1) edge[bend left] node {B / 0.9} (q2)
            ;

        \path [-stealth, thick]
            (q1) edge[bend right] node {$\epsilon$ / 0.1} (q2)
            ;

        \path [-stealth, line width=5pt]
            (q2) edge[bend left] node {C / 0.8} (q3)
            ;

        \path [-stealth, thick]
            (q2) edge[bend right] node {$\epsilon$ / 0.2} (q3)
        ;
    \end{tikzpicture}
    \vspace{2mm}
    
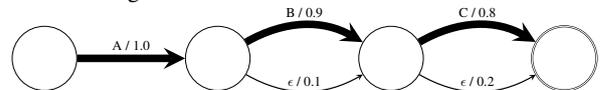
\captionof{figure}{
        Hystoc confidences for $n$-best set ABC, AB, and AC with
        probabilities 0.7, 0.2, and 0.1 respectively.
    }\label{fig:my_label}
\end{minipage}

\section{Results}

The results are presented in Table \ref{tab:results}, in which we indicate the pre-trained feature transform (WavLM Large or Wav2Vec2Large), whether the feature transform was fixed or fine-tuned, the encoder-decoder parts of the ASR model on top of that transform (e.g. Trasformer-Transformer), and the ASR toolkit that was used. The training data indicators refer to Table \ref{tab:data} and for example '(3$^{\dagger}$) $\rightarrow$\textit{gss-only}' indicates that the system was trained as system (3$^{\dagger}$) and then fine-tuned using the \textit{gss-only} data. Development sets were enhanced with the baseline GSS for almost all systems, except for systems (3) and (3$^\dagger$) where 'GSS\_postfilter', i.e.\ GSS with enabled mask-based post-filtering, was used.

Our {\em limited} training data selection presented above (system (1)) in Section~\ref{sec:data} is giving 1.3--2.6\% absolute gain over baseline (0) with significant increase of training speed as the \textit{limited} set contains 10x less data than \textit{baseline} one. Moreover, next 1.2--2.0\% absolute gain is reached with adding `enhanced' Librispeech data for system (2$^{\dagger}$) that was trained with {\em limited+libri} dataset. Additional fine-tuning of this system to the~\textit{gss-only} data is giving further significant 1.2--4.8\% absolute WER reduction resulting in system (2(S)). The system (2(S)) was {\bf submitted} as the single best system to the challenge.

The comparison of systems (2$^{\dagger}$) and (3$^{\dagger}$) shows a drop in ASR performance similar to the one observed in Table \ref{tab:gss_se_dev}, the difference is the reuse use of the CACGMM TF masks on the output of MVDR beamformer as {\em GSS\_postfilter} in (3$^{\dagger}$). Again, the significant drop of WER from (3$^{\dagger}$) to (3) confirms the great importance of fine-tuning with the in-domain data.

The WavLM based systems were found superior to Wav2Vec2 systems, see comparison (4) and (5).
Replacing the CTC loss by a RNN-T loss brought large improvements in WER, as seen in the gap between systems (6) and (7), showing the importance of incorporating a language model for this task. Moreover, when also fine-tuning the CNN encoder in WavLM, and lowering the learning rate from 1e-4 to 1e-5, we observe significant WER improvements between systems (7) and (8). Conditioning WavLM with speaker information didn't bring improvements as seen between systems (8) and (9). More work remains to be done to see if improved results can be achieved by using other conditioning methods like Adaptive Instance Normalization and FiLM layers.


Next, we trained an ASR system with the K2 toolkit using Zipformer-Transducer architecture (system (10)).
Unfortunately, we did not have time to finalize this work for the system presentation at the workshop.
However, this work could be very promising for future, as it significantly outperforms system (1) trained on same data, while both systems are without fine-tuning the WavLM model.

Fusing the best performing systems (F1--F3) did provide modest gains over the individual best systems.
Despite the performance drop observed on the development set for system (3) with {\em GSS\_postfilter}, in the fusion (F3) the system (3) brings complementary information, improving the result compared to fusion (F2).
This phenomenon might be justified by the fact that for some part of the data-set the stronger filtering is actually helpful.

\section{Acknowledgements}
This work has been funded by the EU's Horizon 2020 research and
innovation program under the
Marie Sklodowska-Curie grant agreement No 101007666, the Agency is not
responsible for this results or use that may be made of the
information. This research project has been also supported by the program "Excellence initiative – research university" for AGH University of Krakow.

\bibliographystyle{IEEEtran}

\bibliography{mybib}

\end{document}